# Assessment of the radiological impact of a decommissioning nuclear power plant in Italy

A. Petraglia[a,*], C. Sabbarese[a], M. De Cesare[a], N. De Cesare[b], F. Quinto[a], F. Terrasi[a], A. D'Onofrio[a], P. Steier[c], L. K. Fifield[d], A. M. Esposito[e].

[a.] CIRCE, INNOVA and Department of Environmental Science, Second University of Naples, Via Vivaldi, 43, I-81100 Caserta, Italy.

[b.] CIRCE, INNOVA and Department of Life Science, Second University of Naples, Via Vivaldi, 43, I-81100 Caserta, Italy.

[c.] Vienna Environmental Research Accelerator (VERA), Fakultät für Physik, Universität Wien, Währinger Strasse 17, A-1090 Wien, Austria.

[d.] Department of Nuclear Physics, Research School of Physics and Engineering, The Australian National University, Canberra, ACT 0200, Australia.

[e.] Società Gestione Impianti Nucleari, Garigliano Power Plant, Sessa Aurunca (Ce), Italy.

* Corresponding author. Tel.:+39 0823 274629, Fax.: +39 0823 274605 E-mail : antonio.petraglia@unina2.it

*Abstract*

*The assessment of the radiological impact of a decommissioning Nuclear Power Plant is presented here through the results of an environmental monitoring survey carried out in the area surrounding the Garigliano Power Plant. The levels of radioactivity in soil, water, air and other environmental matrices are shown, in which α, β and γ activity and γ equivalent dose rate are measured. Radioactivity levels of the samples from the Garigliano area are analyzed and then compared to those from a control zone situated more than 100 km away. Moreover, a comparison is made with a previous survey held in 2001. The analyses and comparisons show no significant alteration in the radiological characteristics of the area surroundings the plant, with an overall radioactivity depending mainly from the global fallout and natural sources.*

## 1. Introduction

The measurement and analysis of the radiological impact of a nuclear power plant (NPP) have several points of interest. In particular, environmental radiological surveys are needed to assess contamination levels over the years (Adliene *et al.*, 2006; Lu *et al.*, 2006; Thinova and Trojek, 2009) in order not only to safeguard the health of people and other living organisms, but also to lower the level of risk perception among the population, for which an objective and verifiable scientific survey can be very effective (Hu *et al.*, 2010). This work reports the studies made to assess the radiological impact of the decommissioning procedure of the Garigliano NPP. This plant used a 160 MW boiling water reactor and was active from 1964 to 1979, when it was switched off for maintenance. It was definitively stopped in 1986, when, after a referendum, the procedure for safety storage began. The reactor is currently isolated, the pipes and other component parts are sealed. No fuel is currently in the plant and radioactive intermediate and low level waste is collected in sealed tanks. Preliminary studies and other details regarding the plant and the procedure of decommissioning are reported by (Esposito, 2005).

Here we report about a measuring survey conducted in 2008-2009 in the area surrounding the plant of Garigliano. The results and analysis of both *in situ* and laboratory measurements are shown here and added to studies already carried out in the neighborhood of the Garigliano NPP since 2001 ( Sabbarese *et al.* 2005), so that they may be included in a wider temporal context.

The specific activity of natural and anthropogenic radionuclides was measured in air, ground water, soil and environmental matrices of special interest. In particular we made measurements of
- γ equivalent dose rate in air.

Measurements were made in several stages: after a first assessment, in which the results showed a tendency to increase approaching the volcanic hills near Roccamonfina, new samples were collected in order to better study these trends. Altogether 61 measurements were made.
- γ activity in the air.

The suspended particulate in air was collected by suction on a filter and then analyzed with γ spectrometry. 11 samples were collected.
- Total α and β activity of ground water.

A series of water samples in 6 wells surrounding the NPP area were collected.
- γ activity in soil samples.

The sampling was carried out on undisturbed soil for a total of 53 sampling points.
- γ activity in food samples.

These measurements are used to evaluate the transfer of any pollutants from the soil, air and water sectors to biotic ones and therefore humans. Seventeen samples of vegetables, fruit, river fish, shellfishes (in the sea near the mouth of the Garigliano river) were taken.
- Stratigraphic γ activity in soil samples.

In two selected points γ activity was measured in different soil layers for a total of 10 measures.
The presentation of all results and their discussion are the subject of this paper.

## 2. Material and methods

### 2.1. Sampling

The sampling points were chosen in a circular grid in order to ensure their regular distribution and homogeneity. The area surrounding the plant was divided into circular crowns of one kilometer each, beginning 500 meters from the NPP. The two inner rings were divided into eight angular sectors and the other two in sixteen sectors, for a total of 48 sampling points. The sampling grid is show in Figure 1.

Further 4 sampling points were selected along the side of Roccamonfina hill, east-northeast of the NPP, and 2 points in the Sele plain, in the province of Salerno, geologically similar to the Garigliano plain but geographically far away (130 km) to have a comparison term.

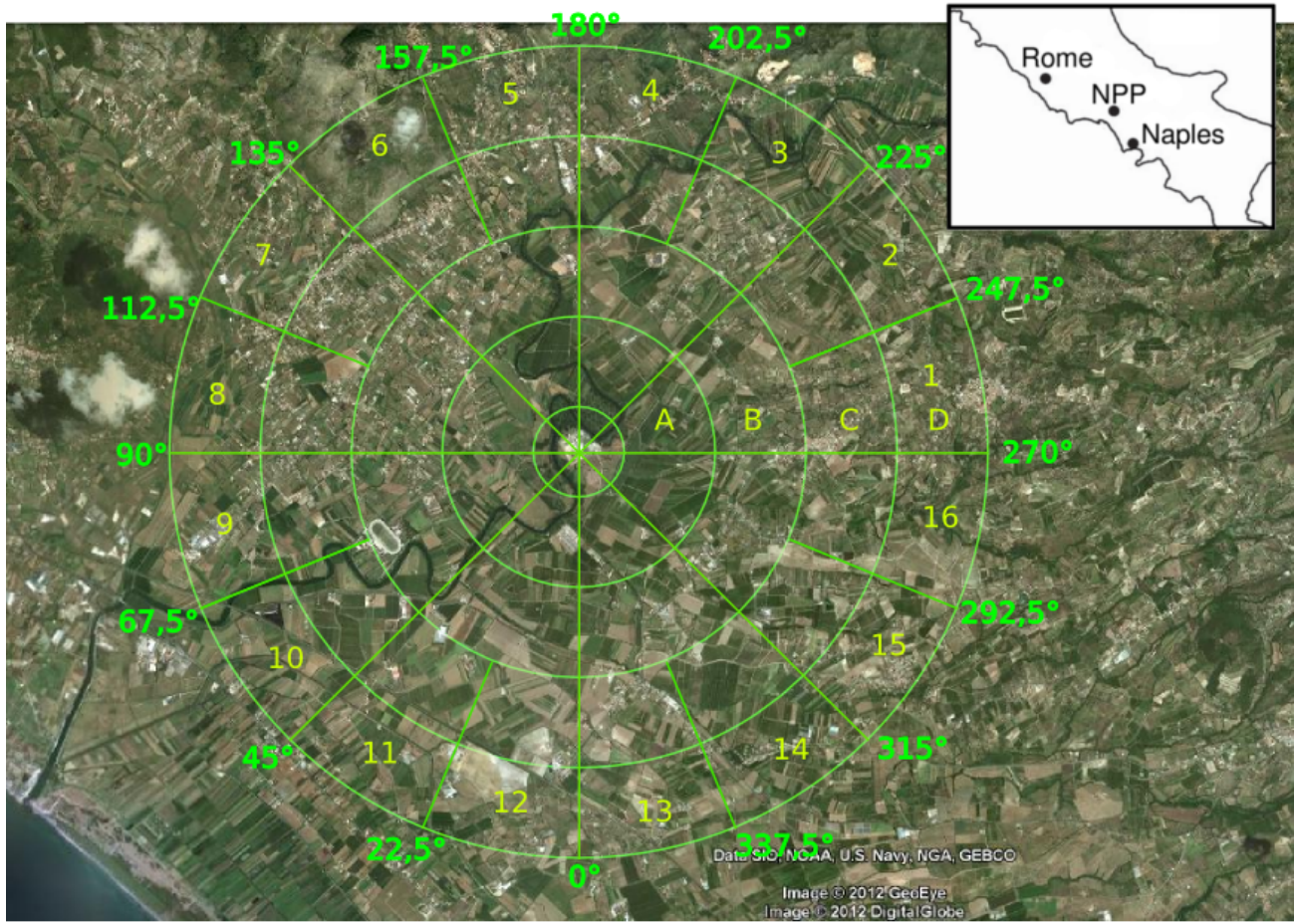

*Figure 1. The sampling grid centered on the Garigliano power plant. North is up. The inset shows the position of the NPP in south/central Italy.*

### 2.2. Sample preparation and measurements

The measurements regarding the γ equivalent dose rate were made with a mobile ionization chamber (FHT 191 N connected to an ESM FHT 6020 by Thermo, energy range: 35 keV-7 MeV), deployed 2 m above the ground. The sampling was extended beyond the selected grid to better understand the spatial distribution of doses going to the east.

The water samples for groundwater α and β activity measurements were collected in six wells in the range of 530-2200 m from the NPP. Samples of 5 dm$^3$ were taken and then dried; the residue was measured for total α (with a Berthold LB770-PC Low-level Planchet Counter) and β activity (with a Wallac Quantulus 1220 ultra-low level liquid scintillation system).

For all the γ measurements the samples were analyzed with a high-resolution germanium hyperpure γ-ray detector (1.9 keV resolution at 1.332 MeV and 70% efficiency) properly shielded; spectra were acquired, displayed and analyzed on computers running Silena Gamma+ and Ortec GammaVision.

For the measurement of the activity in air, suspended particulates were collected by suction on a filter at 11 points on the sampling grid; the total amount of filtered air for each sample varied from 2500 to 3000 l.

For the measurements of the activity in the soil, 20 x 20 x 20 cm$^3$ soil samples were collected on undisturbed ground around the plant; they were then dehydrated, homogenized, sieved and put in Marinelli vessels. A total of 53 points were sampled. Of those, 47 points were in the sampling grid

(one point was not sampled because it was situated in an area where access was difficult). To these, 4 samples were added taken on the slope of the Roccamonfina volcano and 2 control samples in the Sele plain. Two points for stratigraphic measurements were chosen at coordinates E 401327, N 4564206 and E 402636, N 4565181 (UTM coordinates, WGS84 ellipsoid, zone 33T), respectively 4 and 3 km south of the NPP. For the first point (ND12) 4 samples were collected at different stratigraphic levels of 6 cm each; for the other point (NC13), 6 samples of thickness 5 cm have been collected.

For the activity in food, samples of plants (pears, khaki, apples, oranges, lemons, prunes, eggplants, beans, tomatoes, endive, cauliflower, zucchini) and river fish and shellfish from the Garigliano mouth (barbi, mullet, chub, mussels, clams) were collected in spring 2009; they were dehydrated, homogenized, sieved and measured for γ activity with the instrumentation described above. The quantitymass of the prepared samples was varied from 70 g for the vegetables very rich in water, to 400 g for the fish and shellfish.

# 3. Results and discussion

### 3.1. γ equivalent dose rate

The spatial arrangement and values of dose rates, ranging from 0.08 μSv/h (green) to 0.20 μSv/h (red), is shown in Figure 2. The location of the plant is shown in gray, the circles radius is proportional to the measured rates. It is clear that the NPP area shows no patterns, while the higher dose rates were measured far from it, on the slope of the volcano Roccamonfina. In particular, higher values have been measured in areas with rocky outcrops of leucitica tephra produced during past eruptions, very rich in the natural radionuclide $^{40}$K (Lima *et al.*, 2005).

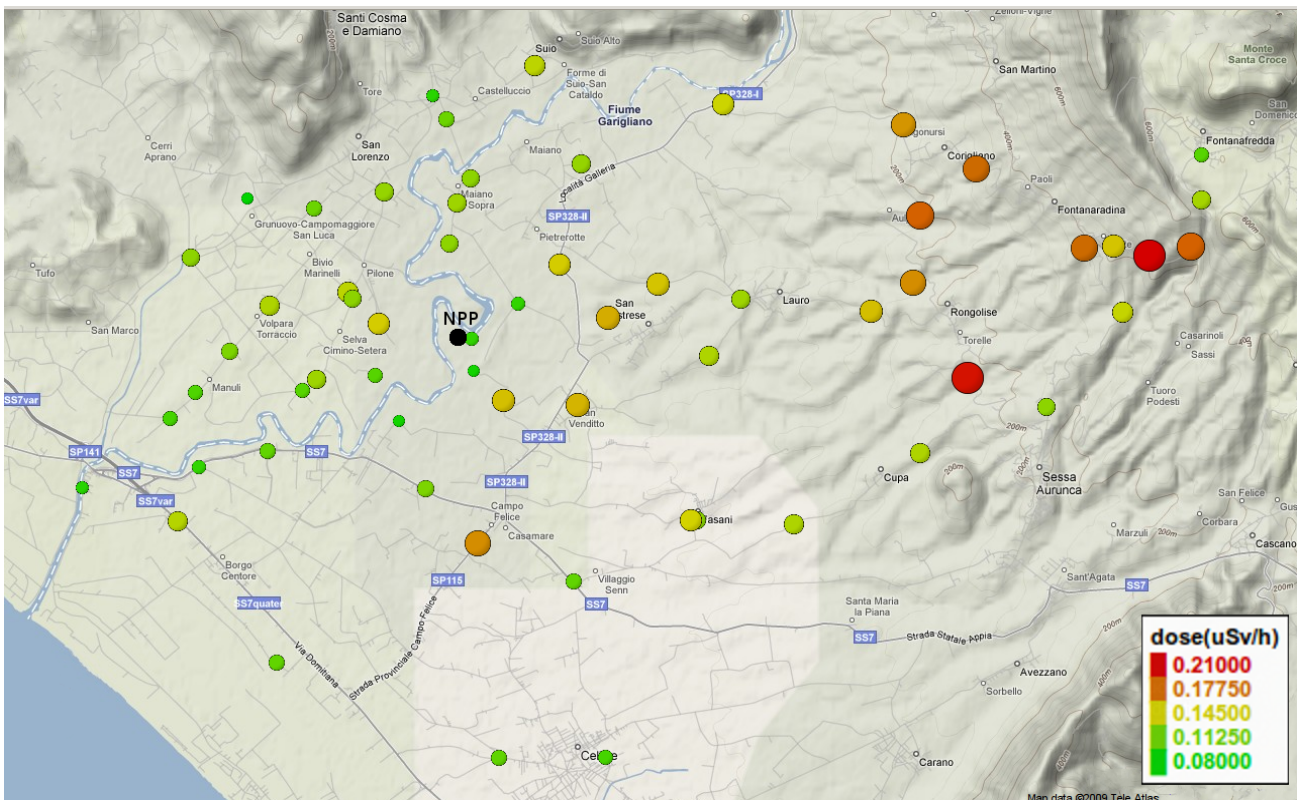

*Figure 2. Spatial arrangement of dose rates. Low values corresponds to smaller green circles, higher values to larger red circles, according to the scale reported in the inset.*

## 3.2. Groundwater α and β activity measurements

The prepared samples, from groundwater collected in neighboring wells, were analyzed for α and β activity. Results are shown in table I.

The total α activity is always below the Minimum Detectable Activity (MDA) for all the samples, indicating low content of α emitting radionuclides. The total β activity values are also shown in Table I. The limits of attention i.e. the "screening values" below which no further analysis is needed, suggested by WHO for drinking water are (WHO, 2008): 0.5 Bq/dm$^3$ for the total α specific activity, and 1.0 Bq/dm$^3$ for the total β specific activity. All the samples were well below these WHO limits.

| UTM coord. (m) | | Total α activity (Bq/dm$^3$) | | Total β activity (Bq/dm$^3$) | |
| --- | --- | --- | --- | --- | --- |
| *easting* | *northing* | **Activity** | *MDA* | **Activity** | *MDA* |
| 404045 | 4568003 | < | 0,0037 | 0,222 | 0,014 |
| 404049 | 4567458 | < | 0,0029 | 0,133 | 0,011 |
| 403867 | 4567162 | < | 0,0034 | 0,072 | 0,012 |
| 403231 | 4566358 | < | 0,0041 | 0,137 | 0,014 |
| 402180 | 4567198 | < | 0,0067 | 0,004 | 0,015 |
| 400889 | 4568698 | < | 0,0020 | 0,058 | 0,012 |

*Table I. Summary of α and β activity measurements in the groundwater samples. Coordinates correspond to UTM zone 33T, WGS84 ellipsoid.*

## 3.3. γ activity in air and soil

The particulate collected in air filters was measured for γ activity with the γ spectroscopy setup described above. A wide range of radionuclides were considered in the analysis, both natural and anthropogenic, but all gave values below the MDA (whose maximum was $2.0 \cdot 10^{-2}$ Bq/m$^3$ for both $^{137}$Cs and $^{60}$Co). Similarly, the γ-ray activities of the 53 soil samples were measured with the same setup, in particular the anthropogenic radionuclides, $^{60}$Co and $^{137}$Cs, were investigated as possible fingerprint of nuclear contamination as well as two natural radioisotopes, $^7$Be and $^{40}$K.

Figure 3 shows the measured activities for the radionuclides under investigation versus distance from the NPP. The main contribution is due to $^{40}$K, followed by $^{137}$Cs with activity values at least an order of magnitude lower. Activity due to $^7$Be was even lower and 34 out of the 51 samples showed $^7$Be activities lower than MDA. No $^{60}$Co activity was detectable in any of the samples (MDA was on average $1.8 \cdot 10^{-3}$ Bq/g for $^7$Be and $1.6 \cdot 10^{-4}$ Bq/g for $^{60}$Co).

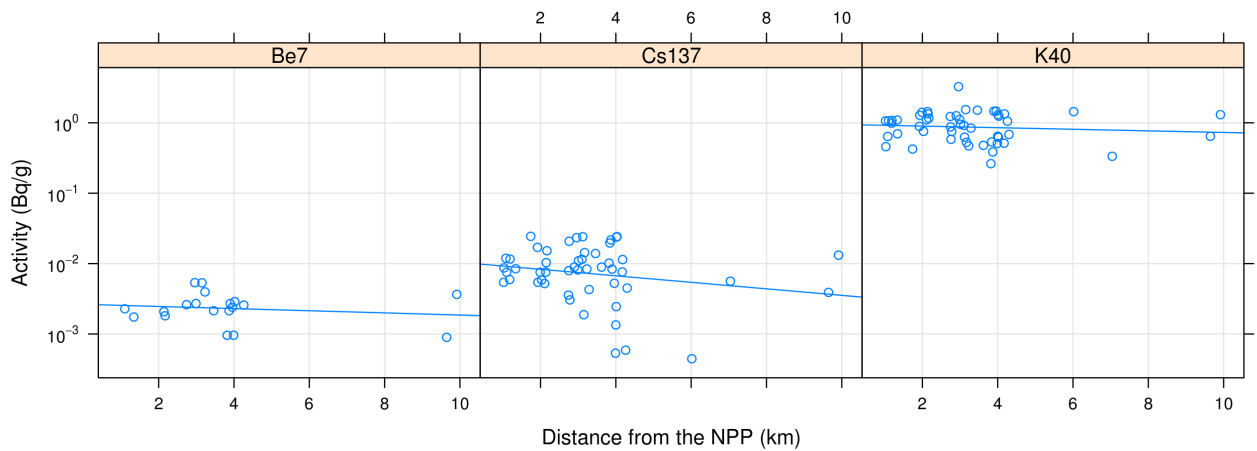

*Figure 3. Measured γ activities in undisturbed soil samples versus distance from the NPP. No $^{60}$Co activity was detectable in any of the samples.*

The analysis of the trends of activity as a function of distance from the NPP together with its linear regression (shown in Figure 3) indicate no correlation between activity and distance. This is confirmed by Table II (calculated by the code R (R, 2009)) which shows the summary of the linear regression for the $^7$Be, $^{40}$K and $^{137}$Cs radionuclides. For each of them there are the estimated value of the slope with its errors and the p-value (marked "Pr (> t)), which indicates the probability of a value less than or equal to the estimated value if the null hypothesis was true. It follows very clearly that the distance from the plant does not influence the activity in a statistically significant manner.

|            | *Slope*  | *standard err.* | *Pr (> t)* |
|------------|----------|-----------------|------------|
| $^7$**Be**     | -3.8E-5  | 1.3E-4          | 0.78       |
| $^{40}$**K**   | -1.3E-2  | 3.8E-1          | 0.73       |
| $^{137}$**Cs** | -3.7E-4  | 5.4E-4          | 0.49       |

*Table II. Summary of the linear regression of activity vs. distance.*

At one point, 3 km south of the NPP (UTM coordinate, WGS84 ellipsoid, zone 33, E 402636, N 4565181), there is a value of activity for $^{40}$K that slightly differs from the average. Further investigations on it are presented below.Figure 4 shows the comparison chart of the values of activity in soil samples in the plains of the Garigliano and Sele rivers.

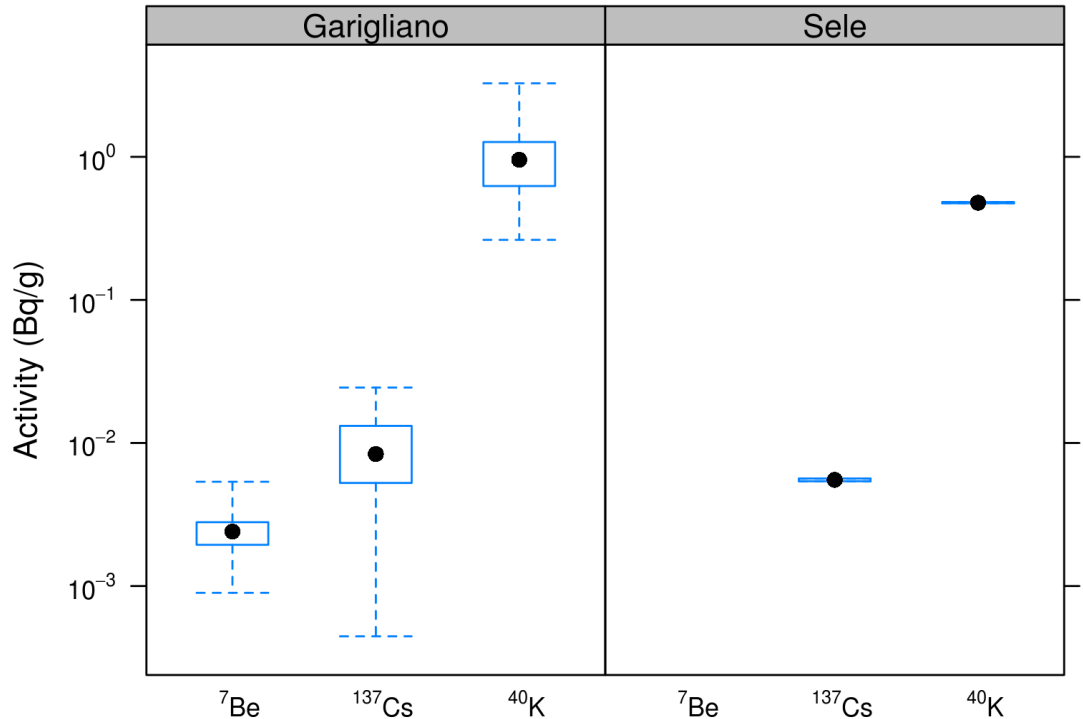

*Figure 4. Comparison chart of the values of γ activity in soil samples in the plains of the Garigliano and Sele rivers. The points are presented in the form of boxplot, with the median (filled circle), the first and third quartiles (box) and range of the data (dashed lines). Missing values denote no detectable radioactivity.*

It is evident that there is good agreement between the two sites for the average values of both the anthropogenic $^{137}$Cs and natural $^{40}$K; the activities of the $^{7}$Be in the Sele plain were both lower than MDA (as were 32 out of 51 measures of the Garigliano plain).

The values of $^{137}$Cs agree with the specific activity corresponding to the fallout levels due to nuclear testing during the Cold War and the Chernobyl accident (Eisenbud and Geisell, 1997; De Cort *et al.*, 1998; Jones *et al.*, 1999; Wallberg *et al.*, 2002; Lu *et al.*, 2006); moreover, expected $^{60}$Co levels due to fallout are consistent with the fact that we measured no detectable activity (Hu *et al.*, 2010 and references above).

From these results we can draw some conclusions:

1. the main contribution to environmental radioactivity is due to potassium ($^{40}$K);
2. there is no correlation between the distance from the NPP and the radioactivity values in the soils for the anthropogenic nuclide $^{137}$Cs;
3. measured values for $^{137}$Cs are in agreement with others reported in the literature and

corresponding to the fallout in Italy;
4. measured values are in agreement with those taken in the control area situated about 130 km away.

**3.4 Measurements of γ activity of food samples and comparison with the environmental survey of 2001.**

The analysis of environmental matrices is necessary to examine the possible contamination around the plant. Most of them showed activities below MDA (whose maximum was $5.7 \cdot 10^{-4}$ Bq/g for $^7$Be, $4.7 \cdot 10^{-5}$ Bq/g for $^{137}$Cs and $6.4 \cdot 10^{-5}$ Bq/g for $^{60}$Co) ; the natural radionuclide $^{40}$K always being the most important contribution, and often the only detectable activity. In Figure 5, on the right column, the γ activity measured in some of these samples is shown.

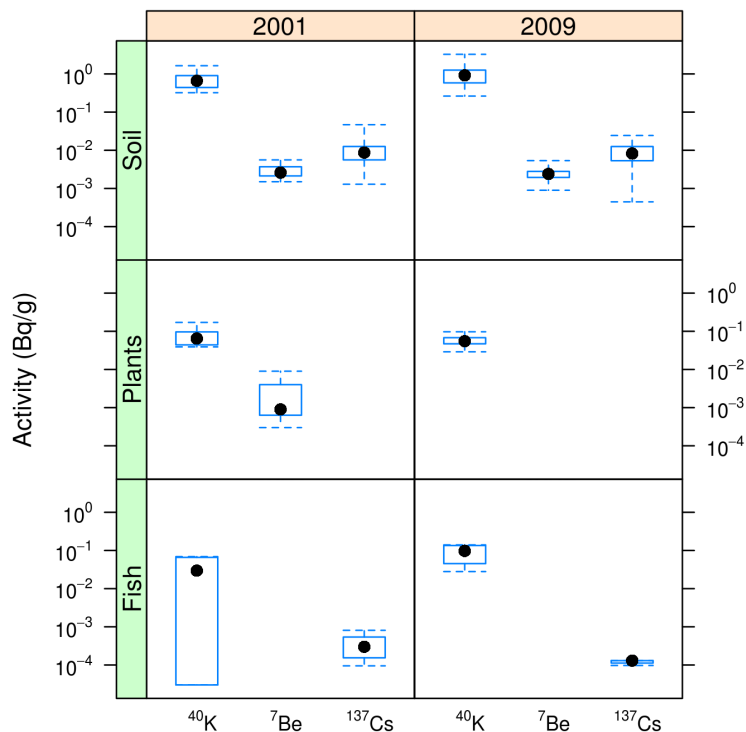

*Figure 5. Synoptic boxplot of the measured γ activity in the environmental samples of the 2001 (left) and 2009 (right) survey. Missing values denote no detectable radioactivity.*

For these measurements it was possible to compare the results with those of a previous survey performed in 2001. In particular it was possible to make a direct comparison of the gamma activity of soil, plants and fish samples, because similar species were taken.

The comparison is presented in Figure 5. We may notice that:
1. there are no substantial differences between the values of the first and second survey;
2. the greater contribution is given by the $^{40}$K (note that the ordinate scale is logarithmic);

**3.5 Activity measurements in stratigraphic samples of soil**

Further analyses have been made for the two locations where activity levels were among the highest. The aim was to verify the previous results and to study the stratification of radionuclides.

The results are shown in Figures 6a, b, c, where the error bars indicate the 3σ intervals (i.e 99.7% of the values fall in the intervals, assuming normal distribution). The average values of $^{40}$K were in agreement with those previously shown in section 2.5 and are uniform, as expected, because it is intrinsic to the soil. Somewhat unexpected, however, is that the $^{137}$Cs activity is also constant with depth, apart from a small increase in the 15-20cm layer of the point NC13, and even the short-lived $^{7}$Be (half-life 53 days) has penetrated to a depth of 15-20 cm at both sites. The $^{7}$Be data may suggest that, despite appearances, the top 15-20 cm of soil has been disturbed over the preceding few months or that several cm of topsoil has been washed in. The $^{137}$Cs data is perhaps less surprising since more than 40 years has passed since deposition, and there is good evidence for bioturbation in the form of ant nests in the near vicinity.

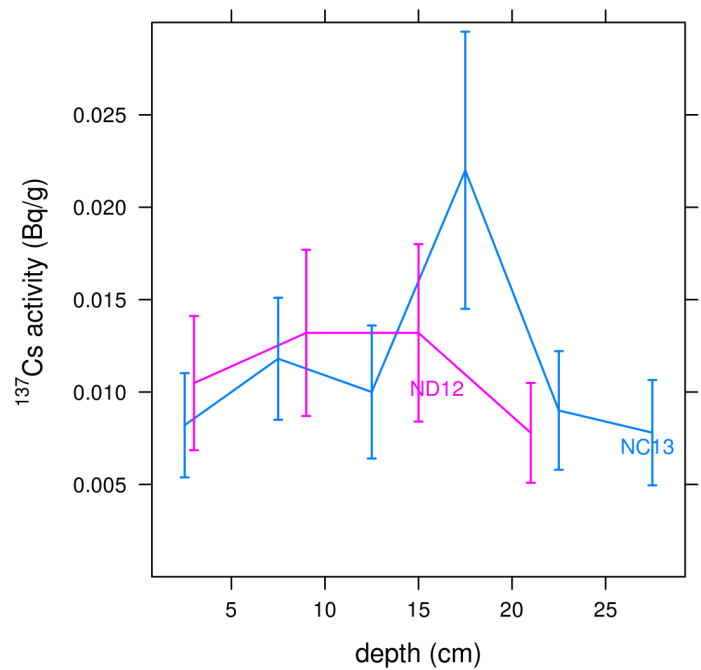

*Figure 6a. Measured γ activity of $^{137}$Cs in stratigraphic samples of soil in two selected point ND12, NC13.*

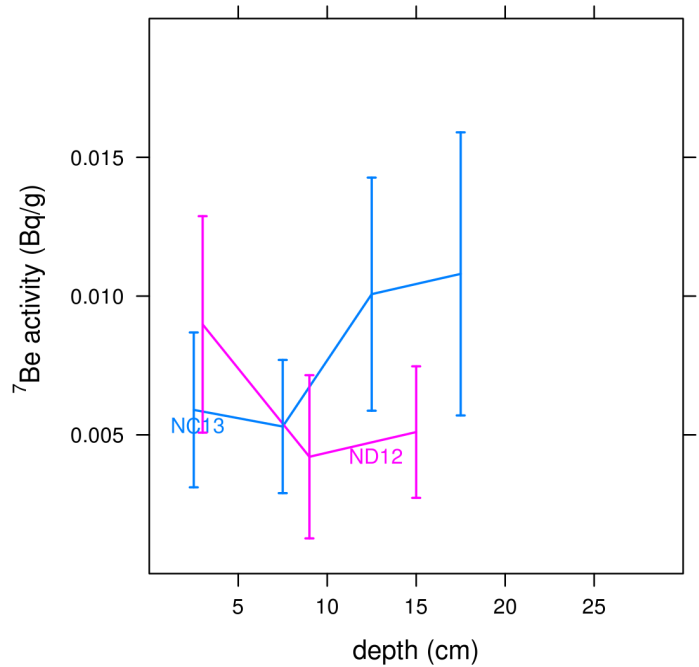

*Figure 6b. Measured γ activity of ⁷Be in stratigraphic samples of soil in two selected point. Missing values denote no detectable radioactivity.*

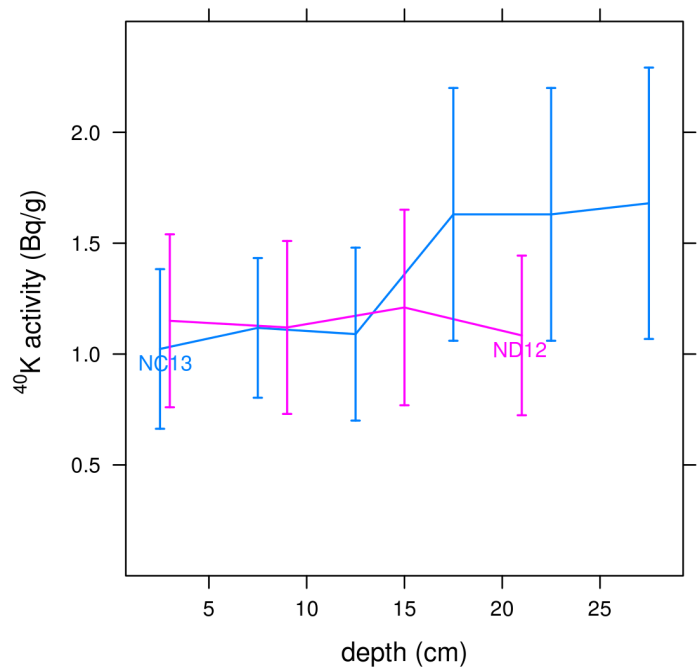

*Figure 6c. Measured γ activity of ⁴⁰K in stratigraphic samples of soil in two selected point.*

It should be reported, here, that during the survey, sediment samples were collected along the river banks for $^{236}U/^{238}U$ isotope ratio measurements. The measurement of nuclear activities and isotopic ratios in environmental samples with AMS (accelerator mass spectrometry) allows to identify the origin of the radionuclide due to the sensitivity of the technique used. These tests represent the state of the art in this field and are carried out for the first time in Italy (De Cesare *et al.*, 2010b). Indeed the ratio of the two uranium isotopes 236 and 238 can address anthropogenic contamination ( Hotchkis *et al.*, 2000). The AMS technique allows measurements to be made on only a few grams of material, so it was also possible to make measurements as a function of depth. The full detail of the sampling and preparation is given in Quinto *et al.*, 2009 together with all the results.

Measurements have also been done on a series of soil samples from around the plant as well as a control sample from the plain of the Sele River; the results and discussion are reported in (De Cesare, 2009; De Cesare *et al.*, 2010a, De Cesare *et al.*, 2012).

The values show little or no contamination due to the plant in agreement with the results reported in this work. Although somewhat elevated levels of $^{236}$U are found in the sediment samples of the drain channel, these are always of the order of the global fallout and much less than in contaminated sites (Berkovits *et al.*, 2000; Boulyga and Heumann, 2006).

## 4. Conclusions

From the results obtained in this work we can draw some conclusions:

- The values of γ activity in soil are dominated by $^{40}$K; the anthropogenic contribution is consistent with fallout from atmospheric nuclear testing during the Cold War years and from the Chernobyl accident, and in agreement with those of a previous survey.
- The same values are consistent with those measured in the alluvial plain of the Sele river, at a considerable distance from NPP.
- No statistically significant correlations between the activities of the anthropogenic radionuclides ($^{60}$Co, $^{137}$Cs) and distance from the NPP have been observed. This also confirms an origin of the latter due to fallout.
- The γ radiation background grows, in a statistically significant manner, approaching the volcanic hill of Roccamonfina, rich in potassium, thus suggesting a natural origin.
- No notable changes occurred between 2001 and 2009 in the content of radioactivity, both natural and anthropogenic, in the area under study.
- The measurements of actinides in sediment samples collected along the river show none or minimal contamination; their values are always at the levels of global fallout or below.

**Acknowledgments**

The technical and administrative staff people of the nuclear power plant of Garigliano are gratefully acknowledged for their hospitality and help. We thanks also E. Cuoco and S. De Francesco for useful discussions.